\documentclass[prd,aps,twocolumn,preprintnumbers, showpacs, nofootinbib,superscriptaddress,notitlepage]{revtex4-1}
\usepackage{amssymb, amsthm}
\usepackage{amsmath}    
\usepackage{mathrsfs}   
\usepackage{graphicx}   
\usepackage{color}      
\usepackage{footnote}   
\usepackage{mathtools}

\usepackage{slashed}    
\usepackage{subfigure}  
\usepackage{hyperref}   

\usepackage{rotating}   
\usepackage{multirow}   
\usepackage[normalem]{ulem} 
\usepackage[utf8]{inputenc}

\newcommand{\beq}{\begin{equation}}
\newcommand{\eeq}{\end{equation}}

\newcommand{\nn}{\nonumber}

\newcommand{\notleftright}{\mathrel{\ooalign{$\leftrightarrow$\cr\hidewidth$/$\hidewidth}}}

\DeclarePairedDelimiter\floor{\lfloor}{\rfloor}

\begin{document}

\preprint{SI-HEP-2021-10}
\preprint{JLAB-THY-21-3341}

\title{Renormalization and Mixing of Staple-Shaped Wilson Line \\ Operators on the Lattice Revisited}

\author{Yao Ji}
\email{yao.ji@uni-siegen.de}
\affiliation{Theoretische Physik 1, Naturwissenschaftlich-Technische Fakult\"at, \\
Universit\"at Siegen, D-57068 Siegen, Germany}

\author{Jian-Hui Zhang}
\email{zhangjianhui@bnu.edu.cn}
\affiliation{Center of Advanced Quantum Studies, Department of Physics, Beijing Normal University, Beijing 100875, China}

\author{Shuai Zhao}
\email{szhao@odu.edu}
\affiliation{Department of Physics, Old Dominion University, Norfolk, VA 23529, USA}
\affiliation{Theory Center, Thomas Jefferson National Accelerator Facility, Newport News, VA 23606, USA}

\author{Ruilin Zhu}
\email{rlzhu@njnu.edu.cn}
\affiliation{Department of Physics and Institute of Theoretical Physics,
Nanjing Normal University, Nanjing, Jiangsu 210023, China}

\begin{abstract}
Transverse-momentum-dependent parton distribution functions and wave functions (TMDPDFs/TMDWFs) can be extracted from lattice calculations of appropriate Euclidean matrix elements of staple-shaped Wilson line operators. We investigate the mixing pattern of such operators under lattice renormalization using symmetry considerations. We perform an analysis for operators with all Dirac structures, which reveals mixings that are not present in one-loop lattice perturbation theory calculations. We also present the relevant one-loop matching in a renormalization scheme that does not introduce extra non-perturbative effects at large distances, both for the TMDPDFs and for the TMDWFs. Our results have the potential to greatly facilitate numerical calculations of TMDPDFs and TMDWFs on the lattice.
\end{abstract}

\maketitle

\section{Introduction}
\label{SEC:Introduction}

 Understanding the transverse structure of hadrons is an important step towards the three-dimensional imaging of hadrons. One of the key quantities that characterizes such transverse structure is the transverse-momentum-dependent parton distribution functions (TMDPDFs), which are a natural generalization of collinear PDFs to incorporate the transverse momentum of partons in the hadron, and provide crucial inputs for describing multi-scale, noninclusive observables at high-energy colliders such as the LHC~\cite{Lin:2020rut}. Currently, our knowledge of TMDPDFs mainly comes from studies of Drell-Yan and semi-inclusive deep-inelastic scattering processes where the transverse momenta of final state particles are measured. QCD factorization theorems allow to relating the relevant experimental observables to TMDPDFs via perturbatively calculable kernels, and thus provide the theoretical basis for extracting TMDPDFs from such observables. In the past, there have been various TMDPDF fittings in the literature~\cite{Bacchetta:2017gcc,Scimemi:2017etj,Bertone:2019nxa,Scimemi:2019cmh,Bacchetta:2019sam,Bacchetta:2020gko}. However, calculating TMDPDFs from first principles has been a challenge, because they are nonperturbative quantities defined in terms of light-cone correlations. 

Early lattice efforts have been focused on extracting certain information of TMDPDFs by studying ratios of suitable correlators~\cite{Hagler:2009mb,Musch:2010ka,Musch:2011er,Engelhardt:2015xja,Yoon:2017qzo}, whereas the full distribution also becomes accessible due to the proposal of large momentum effective theory (LaMET)~\cite{Ji:2013dva,Ji:2014gla,Ji:2020ect} which provides, in principle, a general recipe to calculate light-front (LF) correlations from lattice QCD. In the past few years, there has been rapid progress~\cite{Ji:2014hxa,Ji:2018hvs,Ji:2019ewn,Ji:2020jeb,Zhang:2020dbb,Ebert:2018gzl,Ebert:2019okf,Ebert:2019tvc,Shanahan:2020zxr,Shanahan:2019zcq,Ebert:2020gxr,Vladimirov:2020ofp} on how to extract the quark TMDPDFs from appropriately defined quasi-LF correlations involving staple-shaped Wilson line operators. A viable matching between the quasi-TMDPDF and TMDPDF, with a proper Euclidean construction of soft function for the former, has been established, although either for not fully renormalized quasi-TMDPDFs~\cite{Ji:2019ewn} or in a scheme~\cite{Ebert:2019tvc} that introduces undesired nonperturbative effects at large longitudinal distances. In addition, there have been exploratory lattice studies on the soft function~\cite{Zhang:2020dbb} and the Collins-Soper kernel~\cite{Shanahan:2020zxr,Schlemmer:2021aij} controlling the rapidity evolution of the TMDPDFs, as well as on the potential operator mixings under lattice regularization~\cite{Shanahan:2019zcq,Green:2020xco}.

Another important quantity that encompasses information on the transverse structure of hadrons is the TMD wave functions (TMDWFs) or LF wave functions, from which one can actually obtain all parton densities. They are defined by the same staple-shaped Wilson line operators, and thus the lattice computation follows a similar strategy as that for the TMDPDFs~\cite{Ji:2020ect}. The quasi-TMDWF also enters the calculation of soft function through the TMD factorization of a light-meson form factor at large momentum transfer~\cite{Ji:2020ect,Zhang:2020dbb}.

In this work, we perform a systematic analysis of the mixing pattern of staple-shaped Wilson line operators under lattice regularization using symmetry considerations. Similar analysis has been performed for straight Wilson line operators defining the quark quasi-PDFs in Ref.~\cite{Chen:2017mie}, where the authors analyzed the transformation properties of straight Wilson line operators with various Dirac structures and found the same mixing observed in one-loop lattice perturbation theory calculations for Wilson fermions~\cite{Constantinou:2017sej}. The lattice perturbation theory studies have also been extended to quark quasi-TMDPDFs in Ref.~\cite{Constantinou:2019vyb}, revealing certain mixings among operators with different Dirac structures (see also Ref.~\cite{Green:2020xco}). However, a systematic analysis of the operator mixing pattern from symmetry considerations is still missing. Here we generalize the discussion of Ref.~\cite{Chen:2017mie} to staple-shaped Wilson line operators. The results show mixings that are not present in one-loop lattice perturbation theory calculations. We also discuss the renormalization and matching of quasi-TMDPDFs and -TMDWFs in a scheme where no extra non-perturbative effects are introduced at large distances in the renormalization stage, in the same spirit as the hybrid renormalization~\cite{Ji:2020brr} proposed recently for the quasi-PDFs.

The rest of the paper is organized as follows: In Sec.~\ref{SEC:qTMD}, we give a brief overview of the quasi-TMDPDFs and -TMDWFs in LaMET, both are defined in terms of staple-shaped Wilson line operators along spatial directions. We then discuss in Sec.~\ref{SEC:mixing} the transformation properties of such operators and their mixing pattern under lattice regularization. In Sec.~\ref{SEC:renmat} we discuss the renormalization and matching of quasi-TMDPDFs and -TMDWFs in a scheme following the spirit of hybrid renormalization and give the relevant one-loop matching kernel. Finally we conclude in Sec.~\ref{SEC:conclusion}.

\section{Quasi-TMDPDFs and -TMDWFs in LaMET}
\label{SEC:qTMD}

\begin{figure}[tbp]
\includegraphics[width=0.25\textwidth]{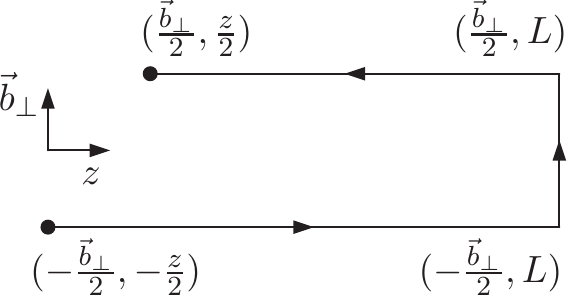}
\caption{Staple-shaped gauge link used to define the quasi-TMDPDF and -TMDWF.}
\label{fig:stapleWL}
\end{figure}

Let us begin with the definition of quasi-TMDPDFs in LaMET with Euclidean metric in four-dimensions~\cite{Ji:2014hxa,Ji:2018hvs,Ebert:2019okf,Ji:2019ewn}
\begin{align}\label{eq:quasi_TMD}
& \tilde f(z ,b_\perp,\mu,P^z) \\
&=\! \lim_{L \rightarrow \infty}  \frac{\langle PS| \bar \psi\big(\frac{\vec z+\vec{b}_\perp}{2}\big)\Gamma{\overline W}(\vec z, \vec{b}_\perp;\vec L)\psi\big(-\!\frac{\vec z+\vec{b}_\perp}{2}\big) |PS\rangle}{\sqrt{Z_E(2L,b_\perp,\mu)}} \ , \nonumber
\end{align}
where we have chosen a symmetric setup to simplify the analysis. $P=(P^0,0,0,P^z)$ is the hadron momentum and $S$ denotes its spin, $\vec L\equiv L  n_z$, $\vec z=z n_z$ with $ n_z=(0,0,0,1)$ being a unit four-vector along the spatial $z$ direction, and $\vec{b}_\perp=(0,b_1,b_2,0)$. The staple-shaped Wilson line takes the following form
\begin{align}\label{eq:staplez}
{\overline W}(\vec z, \vec{b}_\perp;L)&={W_z^\dagger\Big(\vec L+\frac{\vec b_\perp}{2}; \frac{\vec z}{2}-\vec L\Big)}
W_{\perp}\Big(\vec L -\frac{{\vec b}_\perp}{2};\vec b_\perp\Big)\nonumber\\
&\times W_{z}\Big(-\frac{\vec z+\vec{b}_\perp}{2};\vec L+\frac{\vec z}{2}\Big), \nonumber\\
W_{i}(\eta;L)&= {\cal P}{\rm exp}\Big[-ig\int_{0}^{L} dt\, {n}_i\cdot A(\eta^\mu+t n_i^\mu)\Big],
\end{align}
for an illustration see Fig.~\ref{fig:stapleWL}. $\Gamma$ denotes a Dirac matrix. 
$\sqrt{Z_E(2L,b_\perp,\mu,0)}$ is the square root of the vacuum expectation value of a flat rectangular Euclidean Wilson-loop along the $n_z$ direction with length $2L$ and width $b_\perp$:
\begin{align}\label{eq:Z_E}
Z_E(2L,b_\perp,\mu)&=\frac{1}{N_c}{\rm Tr}\langle 0|{W_\perp^\dagger(-\vec\xi_-;-b_\perp)W_z^\dagger(\vec\xi_+;-2L)}\nonumber\\
&{\times W_{\perp}(\vec \xi_-; b_\perp) W_z(-\vec \xi_+;2L)}|0\rangle \, ,
\end{align}
where
\begin{align}
\vec \xi_\pm = L \vec n_z \pm \frac{\vec b_\perp}{2}\, .
\end{align}
In contrast to the usual TMDPDF which contains lightlike separations between quark fields, the quasi-TMDPDF defined above involves spatial separations only. However, the same lightcone physics is projected out when the hadron momentum becomes infinite, as one can unboost the hadron at large momentum and apply the boost operator to the spatial correlator in Eq.~(\ref{eq:quasi_TMD}), yielding the same LF correlator defining the TMDPDF~\cite{Ji:2019ewn,Ji:2020ect}. This is similar to shifting from Schr\"odinger picture to Heisenberg picture in quantum mechanics. Note that the LF correlator in TMDPDF leads to rapidity divergences which require a proper regulator. Given the finite hadron momentum, the quasi-TMDPDF can be viewed in a sense as the definition of TMDPDF with the hadron momentum as a rapidity regulator~\cite{Ji:2019ewn}.

In the above definition, also the length of the longitudinal link is kept finite to regulate the pinch-pole singularity associated with infinitely long Wilson lines~\cite{Ji:2018hvs}. Such link length dependence drops out in the ratio of Eq.~(\ref{eq:quasi_TMD}) so that the final result has a proper $L\to\infty$ limit. The introduction of $Z_E$ also removes additional contributions arising from the transverse gauge link. 
From Eq.~(\ref{eq:quasi_TMD}), the momentum space density is given by the following Fourier transform
\beq \label{eq:TMD-mom}
 \tilde f(x, k_\perp,\mu,\zeta_z) = \int\frac{d\lambda d^2\vec b_\perp}{(2\pi)^3}e^{ix\lambda+i\vec k_\perp\cdot \vec b_\perp}\tilde f(\lambda ,b_\perp,\mu,P^z)\, ,
\eeq
with $\lambda=z P^z$ being the quasi-LF distance, and $\zeta_z=(2xP^z)^2$ is the Collins-Soper scale. The thus defined quasi-TMDPDF depends on two scales, $\mu$ and $\zeta_z$. The dependence on $\mu$ is controlled by the renormalization group equation~\cite{Collins:1981uk,Ji:2004wu}
\beq\label{eq:RG_TMD}
\mu^2\frac{d}{d\mu^2}\ln \tilde f(x, b_\perp,\mu,\zeta_z)=\gamma_F(\alpha_S(\mu)),
\eeq
where $\alpha_S=g^2/(4\pi)$, and $\gamma_F$ is most easily obtained from the anomalous dimension of the quark field in the axial gauge $A^z=0$. In the auxiliary field language~\cite{Dorn:1986dt,Ji:2017oey,Green:2017xeu}, a straight segment of Wilson line can be replaced by the two-point function of an auxiliary heavy quark field, $\gamma_F$ then represents the anomalous dimension of the auxiliary heavy-light quark current. The Wilson line cusp anomalous dimension does not enter because it has been canceled between the numerator and denominator in Eq.~(\ref{eq:quasi_TMD}).

The $\zeta_z$ dependence characterizes how the quasi-TMDPDF changes with momentum or rapidity, and the evolution is controlled by the Collins-Soper equation~\cite{Collins:1981uk,Ji:2014hxa}
\beq\label{eq:CS_TMD}
P^z\frac{d}{dP^z}\ln \tilde f(x, b_\perp, \mu, \zeta_z)=K(b_\perp,\mu)+G(\zeta_z,\mu),
\eeq
where $K(b_\perp,\mu)$ is the Collins-Soper kernel that is independent of the rapidity regularization, while $G(\zeta_z,\mu)$ is a perturbative term existing only in the off-light-cone regularization scheme, its explicit expression at one-loop can be found in Ref.~\cite{Ji:2019ewn}. 

Analogously, one can define the quasi-TMDWF with the same staple-shaped Wilson line operator, but now between the vaccum and a hadron state~\cite{Ji:2020ect}
\begin{align}\label{eq:qTMDWF}
&\tilde\psi(z,b_\perp,\mu,P^z)\\
&=\! \lim_{L \rightarrow \infty}  \frac{\langle 0| \bar \psi\big(\frac{\vec z+\vec{b}_\perp}{2}\big)\Gamma{\overline W}(\vec z, \vec{b}_\perp;\vec L)\psi\big(-\!\frac{\vec z+\vec{b}_\perp}{2}\big) |PS\rangle}{\sqrt{Z_E(2L,b_\perp,\mu)}}. \nonumber
\end{align}
Its scale dependence is controlled by evolution equations similar to Eqs.~(\ref{eq:RG_TMD}) and (\ref{eq:CS_TMD}).

\section{Mixing pattern of staple-shaped Wilson line operators on the lattice}
\label{SEC:mixing}
To calculate the TMDPDFs or TMDWFs, we need to calculate the coordinate space correlation functions defined above on the lattice. A discretized lattice has less symmetry than the continuum, and thus more operator mixings can appear. Moreover, chiral symmetry might be broken after the fermion fields are discretized, leading to additional operator mixings. Nevertheless, the lattice action exhibits important discrete symmetries: parity, time reversal and charge conjugation. Investigating the transformation properties of relevant operators under these symmetries helps to unravel potential mixings that can occur. Such an analysis has been done for straight Wilson line operators defining the quasi-PDFs in Ref.~\cite{Chen:2017mie}. In this section, we extend it to staple-shaped Wilson line operators relevant for the quasi-TMDPDFs and -TMDWFs. 

\subsection{${\cal P}$, ${\cal T}$, ${\cal C}$ and axial transformations}

For the convenience of the reader, we briefly summarize in this subsection the transformation properties of fields under parity (${\cal P}$), time-reversal (${\cal T}$), charge conjugation (${\cal C}$), and the axial transformation. We follow the convention of Ref.~\cite{Chen:2017mie} with the Euclidean spacetime coordinates $(x, y, z, \tau) = (1, 2, 3, 4)$.  Dirac matrices are chosen to be Hermitian: $\gamma_{\mu}^{\dagger}=\gamma_{\mu}$, and $\gamma_5=\gamma_1\gamma_2\gamma_3\gamma_4$.

Since there is no distinction between time and space in Euclidean space, the parity transformation in the $\mu$-direction, denoted by ${\cal P}_{\mu}$ with $\mu\in\{1,2,3,4\}$, can be defined with respect to any direction
\begin{eqnarray}
\psi(x)&\xrightarrow[]{{\cal P}_{\mu}}&
\psi(x)^{{\cal P}_{\mu}}=\gamma_{\mu}\psi(\mathbb{P}_{\mu}(x)),\\
\overline{\psi}(x)&\xrightarrow[]{{\cal P}_{\mu}}&
\overline{\psi}(x)^{{\cal P}_{\mu}}=\overline{\psi}(\mathbb{P}_{\mu}(x))\gamma_{\mu},\\
U_{\nu\not=\mu}(x)&\xrightarrow[]{{\cal P}_{\mu}}&
{U_{\nu}(x)^{{\cal P}_{\mu}}
=U_{{-\nu}}^{\dagger}(\mathbb{P}_{\mu}(x)}
,\\
U_{\mu}(x)&\xrightarrow[]{{\cal P}_{\mu}}&
U_{\mu}(x)^{{\cal P}_{\mu}}=U_{\mu}(\mathbb{P}_{\mu}(x)),
\end{eqnarray}
where $\mathbb{P}_{\mu}(x)$ is the vector $x$ with sign flipped except for the component in the $\mu$-direction. In other words, it is the parity transformation in the $x_\mu$ direction. $U_\mu(x)$ denotes a generic Wilson line along the $\mu$ direction with the starting point at $x$.

Analogously, the time reversal transformation ${\cal T}_{\mu}$ can also be generalized in any direction in Euclidean space
 \begin{eqnarray}
\psi(x)&\xrightarrow[]{{\cal T}_{\mu}}&
\psi(x)^{{\cal T}_{\mu}}=\gamma_{\mu}\gamma_5\psi(\mathbb{T}_{\mu}(x)),\\
\overline{\psi}(x)&\xrightarrow[]{{\cal T}_{\mu}}&
\overline{\psi}(x)^{{\cal T}_{\mu}}=
\overline{\psi}(\mathbb{T}_{\mu}(x))\gamma_5\gamma_{\mu},\\
U_{\mu}(x)&\xrightarrow[]{{\cal T}_{\mu}}&
U_{\mu}(x)^{{\cal T}_{\mu}}=U_{{-}\mu}^{\dagger}(\mathbb{T}_{\mu}({x})
),\\
U_{\nu\not=\mu}(x)&\xrightarrow[]{{\cal T}_{\mu}}&
U_{{\nu}}(x)^{{\cal T}_{\mu}}
=U_{{\nu}}(\mathbb{T}_{\mu}(x)),
\end{eqnarray}
where $\mathbb{T}_{\mu}(x)$ is the vector $x$ with sign flipped only in the $\mu$-direction.

Charge conjugation ${\cal C}$ transforms particles into their antiparticle counterparts. Under charge conjugation, one has
\begin{eqnarray}
\psi(x)&\xrightarrow[]{\cal C}&
\psi(x)^{\cal C}=C^{-1}\overline{\psi}(x)^{\top},\\
\overline{\psi}(x)&\xrightarrow[]{\cal C}&
\overline{\psi}(x)^{\cal C}=-\psi(x)^{\top}C,\\
U_{\mu}(x)&\xrightarrow[]{\cal C}&
U_{\mu}(x)^{\cal C}=U_{\mu}(x)^{\ast}=(U_{\mu}^{\dagger}(x))^{\top},
\end{eqnarray}
with $\top$ denoting the transpose operation, and
\begin{eqnarray}
C\gamma_{\mu}C^{-1}=-\gamma_{\mu}^{\top}, \qquad
C\gamma_5C^{-1}=\gamma_5^{\top}.
\end{eqnarray}

The continuous axial rotation $\cal A$ of the fermion field reads
\begin{align}
\psi(x)&\xrightarrow[]{\cal{A}}\psi'(x)=e^{i\alpha\gamma_5}\psi(x),\nonumber\\
\overline{\psi}(x)&\xrightarrow[]{\cal{A}}\overline{\psi}'(x)
=\overline{\psi}(x)e^{i\alpha\gamma_5}.
\label{eq:chitransf}
\end{align}

\subsection{Operator mixings}
Based on transformation properties of the fields listed above, we can investigate the transformation under discrete symmetries of the following nonlocal operators involving a staple-shaped Wilson line 
\begin{align}\label{eq:O_Gamma}
O_\Gamma(z, \vec b_\perp, L)&=\bar \psi\big(\frac{\vec z+\vec{b}_\perp}{2}\big)\Gamma{\overline W}(\vec z, \vec{b}_\perp;L)\psi\big(-\!\frac{\vec z+\vec{b}_\perp}{2}\big).
\end{align}

Given that the hadron is to be boosted along the $z$-direction, we treat the $z$-direction differently from other directions, as was done in the case of straight Wilson line operators, and categorize the Dirac structure as follows
\begin{eqnarray}
\Gamma\in\{{\bf{1}}, ~\gamma_i, ~\gamma_3, ~\gamma_5, ~i\gamma_i\gamma_5,
~i\gamma_3\gamma_5, ~\sigma_{i3}, ~\epsilon_{ijk}\sigma_{jk}\},
\end{eqnarray}
where $i, j, k\not=3$. 

From the field transformation properties in the previous subsection, one can work out with some effort the transformation properties of $O_{\Gamma}(z, \vec b_\perp, L)$ under discrete symmetries. 
\begin{align}
O_{\Gamma}(z, \vec b_\perp, L)&\xrightarrow[]{{\cal P}_{i\not=3}}
O_{\gamma_i\Gamma\gamma_i}(-z,{-}\vec b_\perp^{{(-)^i}}, -L), \nonumber\\
O_{\Gamma}(z,\vec b_\perp, L)&\xrightarrow[]{{\cal P}_3}
O_{\gamma_3\Gamma\gamma_3}(z,-\vec b_\perp,L),\nonumber\\
O_{\Gamma}(z,\vec b_\perp, L)&\xrightarrow[]{{\cal T}_{i\not=3}}
O_{\gamma_5\gamma_i\Gamma\gamma_i\gamma_5}(z,\vec b_\perp^{{(-)^i}}, L),\nonumber\\
O_{\Gamma}(z,\vec b_\perp, L)&\xrightarrow[]{{\cal T}_3}
O_{\gamma_5\gamma_3\Gamma\gamma_3\gamma_5}(-z,\vec b_\perp, -L),
\end{align}
{where $\vec b_{\perp,j}^{(-)^i}\equiv   (-1)^{\delta_{ij}} \vec b_{\perp,j}$ with $j=1,2$ labeling the component of the transverse vector $\vec b_\perp$. Here no summation is implied over index $j$.}

Under $\cal C$, one has
\beq
O_{\Gamma}(z,\vec b_\perp, L)
\xrightarrow[]{\cal C} O_{(C\Gamma C^{-1})^T}(-z,-\vec b_\perp, L).
\eeq

{We can, therefore, define the following combinations that are eigenstates under discrete symmetries
\begin{align}
{\cal O}^n_\Gamma(z, \vec b_1, L) & = \big\{\big[\big(O_\Gamma(z,\vec b, L)+s_{n0} O_\Gamma(z,b_1,-b_2, L)\big)\notag\\
&\,\,+s_{n1}\left(b_1\mapsto -b_1\right)\big]+s_{n2}\big[z\mapsto -z]\big\}\notag\\
&\quad+s_{n3}\big\{ L\mapsto -L \big\}\, ,
\end{align}
where $s_{ij}=(-1)^{\floor{i/2^j}}$. ${\cal O}^n_\Gamma (z,\vec b,L)$ with $n=1,2,\cdots,16$ is a linear combination of 16 independent operators constructed from freely choosing the sign in front of each argument $z,b_1,b_2,L$. Here $\floor{x}$ is the floor function.
%
{$O_{\Gamma}^n$ therefore form an operator basis with definite ${\cal C}, {\cal P}$, and ${\cal T}$ properties, and 
only operators with the same ${\cal C}, {\cal P}, {\cal T}$ eigenvalues can mix, making the mixing pattern manifest.
Nevertheless, the $O_\Gamma^n$ basis given above is much more complicated than the original one without $-z$, $-b_\perp$, and $-L$ dependence.
Thus, in the following we present the results for the operators $O_n$ defined in Eq.~\eqref{eq:O_Gamma} only, although our analysis is mainly based on the operator basis $O_{\Gamma}^n$.

For the operators in~\eqref{eq:O_Gamma}, Lorentz covariance helps to identify what mixings are forbidden. 
}
For example, the scalar operator $O_1$ does not mix with $O_{\gamma_5}$ and $O_{\gamma_3\gamma_5}$ for the TMDPDF with a staple-shape Wilson line. The same is true for $O_{\gamma_\mu}$ and $O_{\gamma_5}$, $O_{\gamma_3}$ and $O_{\gamma_3\gamma_5}$, 
and etc. These patterns are consistent with the observation in lattice computations \cite{Shanahan:2019zcq,Zhang:2020dbb} taking into account the external off-shell quark state. 
However, it is worth pointing out that the operator basis $O_\Gamma$'s are not complete by themselves, as they are not operators of definite twist and operators of higher twist mix with operators of higher Fock states with additional elementary fields due to QCD equations of motion. In fact, a proper treatment for the mixings between operators observed from one-loop calculation in~\cite{Ebert:2019tvc} requires the introduction of higher Fock states.
This is already evident from operators with straight Wilson line at twist-3 level, see Ref.~\cite{Braun:2021aon} and references therein.
Lorentz symmetry allows certain mixings to be identified as mixings with higher Fock states which are statistically suppressed in general. This explains the smallness of mixings between certain Dirac structures observed in~\cite{Shanahan:2019zcq}. A thorough investigation including operators of higher Fock state is beyond the scope of the present paper, and left to future work.

On the other hand, we find that $O_\Gamma$ mixes in general with $O_{\gamma_3\Gamma}$ for arbitrary Dirac structure $\Gamma$, provided that a fermion action that does not preserve chiral symmetry is used.  
This is in contrast to operators with straight Wilson line, where such mixings do not occur for a specific set of $\Gamma$, e.g., $\Gamma=\gamma_4$ in the unpolarized case. The above mixing pattern contains the mixings observed in lattice perturbation theory calculations ~\cite{Constantinou:2019vyb}, but also contains mixings that are not present in such calculations. Note that the definition of quasi-TMDPDFs or -TMDWFs also involves a factor of $\sqrt {Z_E}$, but it does not change the mixing pattern discussed above, as it is a common factor for operators with all Dirac structures and depends only on the length $b_\perp$ and $L$.

\section{Renormalization and matching of quasi-TMDPDFs and -TMDWFs in a simple scheme}
\label{SEC:renmat}
In this section, we discuss the renormalization and matching of quasi-TMDPDFs and -TMDWFs in a simple scheme that does not introduce extra non-perturbative effects at large distances, following the same spirit as the hybrid renormalization scheme introduced in Ref.~\cite{Ji:2020brr}. 

Using the auxiliary field formalism~\cite{Dorn:1986dt}, the straight Wilson line operators have been shown to be multiplicatively renormalized~\cite{Ji:2017oey,Green:2017xeu,Ishikawa:2017faj}. The same can be shown to be true for the staple-shaped operators, with the renormalization factors eliminating the power divergences associated with the Wilson line, the cusp divergences as well as the endpoint divergences~\cite{Ebert:2019tvc}. Thus, the renormalization of the quasi-TMDPDFs and quasi-TMDWFs can be carried out in analogy with that of the quasi-PDFs, and for the latter a commonly used renormalization scheme is the regularization-independent momentum subtraction (RI/MOM) scheme (or its variation $\rm{RI'/MOM}$)~\cite{Constantinou:2017sej,Stewart:2017tvs,Alexandrou:2017huk,Chen:2017mzz} or the ratio scheme~\cite{Radyushkin:2017cyf,Orginos:2017kos,Braun:2018brg,Li:2020xml}. These schemes have the advantage of avoiding certain discretization effects at short distances. The $\rm{RI'/MOM}$ renormalization for the quasi-TMDPDFs has been discussed in the literature~\cite{Constantinou:2019vyb,Ebert:2019tvc,Shanahan:2019zcq}. However, as pointed out in Ref.~\cite{Ji:2020brr}, both the RI/MOM and the ratio schemes introduce undesired non-perturbative effects at large $z$ in the renormalization stage, and thus become unreliable at large distances. This can be clearly seen in a recent analysis of data at multiple lattice spacings in Ref.~\cite{Huo:2021rpe}. In contrast, the Wilson line mass subtraction scheme~\cite{Chen:2016fxx,Ishikawa:2017faj} does not have this issue. Based on this, an alternative renormalization strategy, the hybrid renormalization, has been proposed in Ref.~\cite{Ji:2020brr}, which utilizes the advantages of different schemes at short and long distances. Another issue with the RI/MOM scheme is that it involves off-shell external states, which bring in a lot of complications when going to higher orders in perturbation theory~\cite{Chen:2020ody} or dealing with gauge particles such as gluons~\cite{Zhang:2018diq,Wang:2019tgg}. This can be avoided if one chooses physical matrix elements for renormalization.

The discussion above indicates that for the quasi-TMDPDFs or -TMDWFs, one shall also switch to a more reliable renormalization scheme such as the hybrid scheme. Fortunately, the lattice study of quasi-TMDPDFs/-TMDWFs is focused on non-perturbative or large $b_\perp$ region (for small $b_\perp$ the TMDPDFs can be studied through a factorization into integrated PDFs, similar factorization is expected also for the TMDWFs). Thus, even at small longitudinal separation $z$ one does not need to worry about the discretization effects plaguing the quasi-PDFs. As a result, we can perform the renormalization in a simple manner by removing the logarithmic and linear divergences separately, in the same spirit as the Wilson line mass subtraction scheme~\cite{Chen:2016fxx,Ishikawa:2017faj}. 

Bearing in mind the mixing discussed in the previous section, we can write down the following renormalization 
\begin{align}\label{eq:renorm}
\begin{pmatrix}
 {\bar O}^B_{\Gamma} \\  {\bar O}^B_{\Gamma'}
\end{pmatrix}
&=
{\cal Z}  
\begin{pmatrix}
 {\bar O}_{\Gamma}^R \\  {\bar O}_{\Gamma'}^R
\end{pmatrix}
=
\begin{pmatrix}
{\cal Z}_{\Gamma\Gamma} & {\cal Z}_{\Gamma\Gamma'}\\ {\cal Z}_{\Gamma'\Gamma} & {\cal Z}_{\Gamma'\Gamma'}
\end{pmatrix}
\begin{pmatrix}
 {\bar O}_{\Gamma}^R \\  {\bar O}_{\Gamma'}^R
\end{pmatrix}
+{\rm{h.f.}}
\nn\\
&=
Z\begin{pmatrix}
{Z}_{\Gamma\Gamma} & {Z}_{\Gamma\Gamma'}\\ {Z}_{\Gamma'\Gamma} & {Z}_{\Gamma'\Gamma'}
\end{pmatrix}
\begin{pmatrix}
 {\bar O}_{\Gamma}^R \\  {\bar O}_{\Gamma'}^R
\end{pmatrix}+{\rm{h.f.}}\ ,
\end{align}
where {${\rm h.f.}$ stands for higher Fock states which are an integral part of a complete operator basis.
They are generated by terms that are proportional to the external quark momenta should an external quark state in momentum space is used in loop calculations.
The investigation of their contributions is beyond the scope of this paper and left for future work.} The superscript $B$ and $R$ denotes bare and renormalized operators, respectively, $\Gamma'=\gamma_3\Gamma$, and we assume the factor $\sqrt {Z_E}$ has been divided in all $\bar O_\Gamma$'s, and ignore the arguments for notational simplicity. In the second row, we have separated an overall renormalization factor $Z$ because the multiplcative renormalization is independent of the Dirac structure involved in the operator~\cite{Ji:2017oey,Musch:2010ka}. Moreover, the linear and cusp divergences associated with the Wilson line are canceled by $\sqrt {Z_E}$, thus one only needs to renormalize the remaining logarithmic divergences at the endpoints of the operator. If the operator mixing is absent, there is an easy way to achieve this. One can calculate the straight line operator matrix elements at two different distances $z_0$ and ${z_0}/2$ (with $z_0$ being in the perturbative region) and form the ratio which effectively removes the factorized linear divergence ${\rm e}^{\delta m |z|}$ from the self-energy of the Wilson-line while still retains the required logarithmic divergence,
\begin{align}\label{eq:renormZ}
Z&=\frac{\tilde h_\Gamma^2(z_0/2, P^z)}{\tilde h_\Gamma(z_0, P^z)}\, ,
\end{align}
where $\tilde h_\Gamma$ can be chosen, e.g., as the zero-momentum hadron matrix element used in the ratio scheme or the off-shell quark matrix element used in the RI/MOM scheme, and $P^z$ is not necessarily the same as the momentum used in calculating the quasi-TMDPDF matrix element. For example, in the unpolarized case, one can use
\begin{align}
&\tilde h_{\gamma^t}(z_0,P^z=0)\\
&=\frac{1}{2P^t}\langle P^z=0|\bar\psi(z_0)\gamma^t W_z(z_0,0)\psi(0)|P^z=0\rangle, \nonumber
\end{align}
where $P$ denotes the hardon momentum, or
\begin{align}
&\tilde h_{\gamma^t}(z_0,p^z=0)\nn\\
&=\left.\frac{\sum_s\langle p,s|O_{\gamma^t}(z_0)|p,s\rangle}{\sum_s\langle p,s|O_{\gamma^t}(z_0)|p,s\rangle_{\rm tree}}\right|_{\tiny{p^2=-\mu_R^2,\,p^z=0}}\, ,
\end{align}
with $p,s$ being the off-shell quark momentum and spin, respectively.
Note that although the RI/MOM matrix element exhibits a non-universal linear divergence behavior depending on the lattice action used~\cite{Huo:2021rpe}, it is still allowed here because in $Z$ all such linear divergences cancel out by construction. In this way, what is left in $Z$ is just the endpoint renormalization factors associated with some perturbative corrections.

However, in the presence of operator mixing, one needs to determine the mixing matrix, which requires calculating certain quasi-TMDPDF matrix elements. One can choose, for example, the hadron matrix element of the quasi-TMDPDF operator or the RI/MOM renormalization factor, but at perturbative $z$ and $b_\perp$, so that extra non-perturbative effects are avoided at the renormalization stage. 

The first option of determining the mixing matrix entries is to follow the calculation of renormalization factors in the presence of mixing in the RI/MOM scheme~\cite{Constantinou:2017sej}. 
Since all $z$- and $b_\perp$-dependent UV divergences have been canceled in $\bar O_\Gamma$, we prefer to choose renormalization conditions such that $Z_{ij}$ are independent of $z$ and $b_\perp$. In other words, we can still apply the RI/MOM renormalization conditions similar to those in~\cite{Constantinou:2019vyb,Shanahan:2019zcq}, but only at a given perturbative $z$ and $b_\perp$, and use the results for the renormalization of correlators at all distances. 
In other words, we may require
\begin{align}\label{eq:RIMOMnew}
&Z_q^{-1}(p) {\cal {\bar Z}}_{\Gamma\Gamma'}{\rm Tr}[\Lambda_{{\bar O}_\Gamma}(p)\Gamma']|_{\tiny{z=z_0,b_\perp=b_{\perp 0},p^\mu=p_0^\mu}} \nn\\
&={\rm Tr}[\Lambda_{{\bar O}_\Gamma}^{\rm tree}(p)\Gamma']|_{\tiny{z=z_0,b_\perp=b_{\perp 0},p^\mu=p_0^\mu}},
\end{align}
where ${\cal {\bar Z}}={\cal Z}^{-1}$, $z_0, b_{\perp0}$ are chosen within the perburbative region, $p_0^\mu=(p_0,0,0,0)$, $\Lambda_{{\bar O}_\Gamma}$ is the amputated Green's function of the operator ${\bar O}_\Gamma$ in an off-shell quark state, and the superscript ``tree" denotes its tree-level value. $Z_q$ is the quark wave function renormalization factor determined as
\beq
Z_q(p)=\frac{1}{12}{\rm Tr}[S^{-1}(p)S^{\rm tree}(p)],
\eeq
with $S(p)$ and $S^{\rm tree}(p)$ denoting the quark propagator and its tree-level value, respectively. 

From Eq.~(\ref{eq:renorm}) and the renormalization factors calculated in Eq.~(\ref{eq:RIMOMnew}), one obtains the renormalized quasi-TMDPDF in the RI/MOM scheme
\begin{align}
\hspace{-1em}{\tilde f}_R(\Gamma,z, b_\perp)&=
{{\cal {\bar Z}}}_{\Gamma\Gamma}\tilde f_B(\Gamma,z,b_\perp)+{{\cal {\bar Z}}}_{\Gamma\Gamma'}\tilde f_B(\Gamma',z,b_\perp),
\end{align}
which can be converted to the $\overline{\rm MS}$ scheme by applying a conversion factor
\begin{align}
{\tilde f}_R^{\overline{\rm MS}}(\Gamma,z,b_\perp)={\tilde C}{\tilde f}_R(\Gamma,z,b_\perp).
\end{align}
The conversion factor in general can take a diagonal form~\cite{Constantinou:2019vyb} or a non-diagonal form~\cite{Ebert:2019tvc}, depending on the choice of projectors. Here we choose to define a diagonal conversion factor which reads for $\Gamma=\gamma^\mu$
\begin{align}
\tilde C_\Gamma&=1+\Big[\frac{1}{6}{\cal V}_\Gamma^{\mu\mu}(\mu,p_0)-Z_q^{(1)}\Big],
\end{align}
where $V^{\mu\mu}_\Gamma$ has been calculated in Ref.~\cite{Ebert:2019tvc}, and no summation is implied over the index $\mu$.

Similar renormalization conditions can be constructed if the quasi-TMDPDF matrix elements of hadrons are used. There are different ways to choose such conditions. We choose to determine the renormalization factors by requiring that the renormalized quasi-TMDPDF matrix element be equal to the tree-level value at given perturbative $z$ and $b_\perp$. To avoid potential collinear divergences in the perturbative result and thus in the renormalization factor, we can set $z=0$ or $P^z=0$. In the following, we illustrate this procedure by taking $\Gamma=\gamma^\mu$ (with $\Gamma'=\gamma_3\gamma^\mu$) as an example, we require that the renormalization factors satisfy
\begin{align}
&\langle P|{\bar O}_\Gamma^B|P\rangle_{\rm tree}=\big[{\cal {\bar Z}}_{\Gamma\Gamma}\langle P|{\bar O}_\Gamma^B|P\rangle\nn\\
&+{\cal {\bar Z}}_{\Gamma\Gamma'}\langle P|{\bar O}_{\Gamma'}^B|P\rangle\big]\big|_{z=z_0,b_\perp=b_{\perp 0},P^\mu=p_0^\mu},
\end{align}
where $P_0^\mu=(p^0,0,0,p^z)$. The corresponding conversion factor can be computed with dimensional regularization in the continuum, which also takes a diagonal form. For $z_0=0$, we have
\begin{align}
{\tilde C}_\Gamma=1-\frac{\alpha_S C_F}{2\pi}\Big(\frac{1}{2}L_b^2-\frac{3}{2}L_b+L_b L_z+\frac{1}{2}L_z^2-L_z+\frac{3}{2}\Big),
\end{align}
with $L_b=\ln(b_\perp^2\mu^2 e^{2\gamma_E}/4)$, $L_z=\ln(4p_z^2/\mu^2)$. While for $p^z=0$, we have
\begin{align}
{\tilde C}_\Gamma=1+\frac{\alpha_S C_F}{4\pi}\Big(L_b+4\ln\frac{b_\perp^2+z^2}{b_\perp^2}-4\frac{z}{b_\perp}\arctan\frac{z}{b_\perp}+1\Big).
\end{align}

With the renormalization and conversion factors above, we can convert the renormalized quasi-TMDPDF to the $\overline{\rm MS}$ scheme, and then match it to the TMDPDF. Following Refs.~\cite{Ji:2019ewn,Ji:2020ect}, the connection between the $\overline{\rm MS}$ scheme quasi-TMDPDF $\tilde f$ and TMDPDF $f^{\rm TMD}$ takes the following form,
\begin{align}\label{eq:quasiTMDmatch}
&f^{\rm TMD}(x,b_\perp,\mu,\zeta)=C\left(\frac{\zeta_z}{\mu^2}\right)\\ &\times e^{-\frac{1}{2}\ln (\frac{\zeta_z}{\zeta})K(b_\perp,\mu)}
{\tilde f}(x,b_\perp,\mu,\zeta_z)S_r^{\frac{1}{2}}(b_\perp,\mu)+...\ ,\nn
\end{align}
where $C(\zeta_z/\mu^2)$ is a perturbative matching kernel and its explicit expression to $O(\alpha_S)$ can be found in Ref.~\cite{Ji:2020ect}. The exponential term contains the Collins-Soper evolution kernel which can be computed from the ratio of $\tilde f$'s at different rapidity scales. $\zeta$ corresponds to the Collins-Soper scale characterizd by the full hadron momentum. $S_r$ is the so-called reduced soft factor. The omitted terms are power corrections of order ${\cal O}\left(\Lambda_{\rm QCD}^2/\zeta_z,M^2/(P^z)^2,1/(b^2_\perp\zeta_z) \right)$. Similar matching relations have also been discussed in Refs.~\cite{Constantinou:2019vyb,Ebert:2019tvc}.

For completeness, let us also summarize here how the remaining terms in Eq.~(\ref{eq:quasiTMDmatch}) can be calculated. The Collins-Soper kernel can be extracted by forming the ratio of quasi-TMDPDFs at two different $\zeta_z$'s~\cite{Ebert:2019tvc}
\begin{align}
\frac{\tilde f_R(x, b_\perp, \mu, \zeta_{z})}{\tilde f_R(x, b_\perp, \mu, \zeta'_{z})}=\frac{C(\frac{\zeta'_z}{\mu^2})}{C(\frac{\zeta_z}{\mu^2})}\Big(\frac{\zeta_z}{\zeta'_z}\Big)^{K(b_\perp,\mu)/2}\, .
\end{align}

As for the soft function $S_r(b_\perp,\mu)$, it 
is calculable from the form factor of a pseudoscalar light-meson state with quark content $\bar\psi\eta$ defined as~\cite{Ji:2020ect,Zhang:2020dbb}
\begin{align}
F(b_\perp, P, P', \mu)=\langle P'| \bar\eta(\vec b_\perp)\Gamma'\eta(\vec b_\perp)\bar\psi(0)\Gamma\psi(0)|P\rangle,
\end{align}
where $P,P'$ are two large momenta approaching opposite light-like directions. Making use of the quasi-TMDWFs $\tilde\psi_{\bar q q}$ in Eq.~\eqref{eq:qTMDWF} with light quark state $q$, the form factor takes the following factorized form,
\begin{align}
&F(b_\perp, P, P', \mu)=\nn\\
&\int dx\, dx'\, H(x, x'){\tilde\psi}^\dagger_{\bar q q}(x', b_\perp){\tilde\psi}_{\bar q q}(x, b_\perp)S_r(b_\perp, \mu),
\end{align}
where the perturbative matching kernel up to one-loop has been given in Ref.~\cite{Ji:2020ect}.

The non-perturbative renormalization and conversion factors also apply to the quasi-TMDWFs. After converting to the $\overline{\rm MS}$ scheme quasi-TMDWFs, one can use the matching derived in Ref.~\cite{Ji:2020ect} to obtain the TMDWFs, where the matching relation takes the same form as Eq.~(\ref{eq:quasiTMDmatch}) with a different matching kernel, whose result at one-loop has also been given in Ref.~\cite{Ji:2020ect}.

\section{Conclusion and outlook}
\label{SEC:conclusion}
In this paper, we have investigated the mixing patterns of staple-shaped Wilson line operators defining the quasi-TMDPDFs and -TMDWFs under lattice regularization using symmetry considerations. Our analysis shows that for non-chiral fermions the mixing with other Dirac structures is generally allowed, except for certain specific cases such as $O_1 \notleftright O_{\gamma_5}$, etc. 
There is no Dirac structure, however, that is \textit{completely} free from mixing with other structures. 
{To be more specific, all $\Gamma$s mix with $\gamma^3\Gamma$.}
This is in contrast with the case of straight Wilson line operators, where for certain choice of $\Gamma$, no mixing occurs. 
{We emphasize that we have excluded operator mixings with higher Fock states, which in itself is not self-consistent as the evolution of higher-twist two-particle TMDPDFs are not autonomous. A complete treatment is, however, beyond the scope of this paper and calls for further investigations.}
We have also discussed the renormalization of quasi-TMDPDFs and -TMDWFs in a simple scheme that does not introduce extra non-perturbative effects at large distances, and presented the relevant one-loop matching. The results will facilitate the numerical calculation of TMDPDFs and TMDWFs on the lattice.

It is worth pointing out that the operator mixing analysis presented here is based on transformation properties of the relevant operators under discrete symmetries, chiral symmetry and Euclidean symmetry only. For a more thorough analysis of the operator mixing pattern, it might be convenient to start from the auxiliary field formalism and replace the nonlocal operators with local ones in the framework of lattice field theory, and study the mixing of the former from that of the latter. This could be important not only for the staple-shaped Wilson line operators but also for the ones with a straight line, given that a different linear divergence behavior has been observed on lattice in the RI/MOM matrix elements from that of hadron matrix elements, which needs to be understood.

\vspace{2em}

\acknowledgments

We thank Yizhuang Liu, Wei Wang, Yibo Yang and Yong Zhao for valuable discussions. YJ is supported by the DFG grant SFB TRR 257. JHZ is supported in part by National Natural Science Foundation of China under grant No. 11975051, No. 12061131006, and by the Fundamental Research Funds for the Central Universities. SZ is supported by Jefferson Science Associates, LLC under  U.S. DOE Contract \#DE-AC05-06OR23177 and by U.S. DOE Grant \#DE-FG02-97ER41028. RZ is supported in part by National Natural Science Foundation of China under grant No. 12075124.

\bibliographystyle{apsrev}
\bibliography{opmx}

\end{document}